\documentclass[aps,prl,reprint,twocolumn,showpacs,superscriptaddress]{revtex4}

\usepackage[dvips]{graphicx}
\usepackage{dcolumn}
\usepackage{bm}
\usepackage{xspace}
\usepackage{multirow}
\usepackage{natbib}
\usepackage{color}

\begin{document}

\preprint{}
\title{Electronic Structure of the Topological Insulator Bi$_2$Se$_3$ Using
    Angle-Resolved Photoemission Spectroscopy: Evidence for a Nearly Full
    Surface Spin Polarization}

\author{Z.-H. Pan}
\affiliation{Condensed Matter Physics and Materials Science Department, Brookhaven National Lab, Upton, New York 11973}
\author{E. Vescovo}
\affiliation{National Synchrotron Light Source, Brookhaven National Lab, Upton, New York 11973}
\author{A. V. Fedorov}
\affiliation{Advanced Light Source, Lawrence Berkeley National Laboratory, Berkeley, California 94720}
\author{D. Gardner}
\author{Y.S. Lee}
\affiliation{Department of Physics, Massachusetts Institute of Technology, Cambridge, Massachusetts 02139}
\author{S. Chu}
\affiliation{Center for Materials Science and Engineering, Massachusetts Institute of Technology, Cambridge, Massachusetts 02139}
\author{G. D. Gu}
\author{T. Valla}
\email{valla@bnl.gov}
\affiliation{Condensed Matter Physics and Materials Science Department, Brookhaven National Lab, Upton, New York 11973}

\date{\today}

\begin{abstract}
We performed high-resolution spin- and angle-resolved photoemission spectroscopy studies of the electronic structure and the spin texture on the surface of Bi$_2$Se$_3$, a model topological insulator. By tuning the photon energy, we found that the topological surface state is well separated from the bulk states in the vicinity of $k_z=Z$ plane of the bulk Brillouin zone. The spin-resolved measurements in that region indicate a very high degree of spin polarization of the surface state, $\sim 0.75$, much higher than previously reported. Our results demonstrate that the topological surface state on Bi$_2$Se$_3$ is highly spin polarized and that the dominant factors limiting the polarization are mainly extrinsic.  
\end{abstract}
\vspace{1.0cm}


\maketitle
\pagebreak
Topological insulators (TIs) represent a new quantum state of matter, distinguished from conventional insulators by a Z$_2$ topological invariant \cite{Kane2005,Kane2005a}.
Similar to conventional insulators, TIs are characterized by an enery gap between the occupied and unoccupied states in the bulk band structure. However, at surfaces of a TI, gapless topological surface states (TSS) always exist. These TSSs show many exotic properties, making them very attractive for possible spintronic and quantum computing applications \cite{Fu2008,Garate2010}. 
The existence of TIs, initially predicted on theoretical grounds \cite{Fu2007,Zhang2009}, has recently been confirmed in angle-resolved photoemission spectroscopy (ARPES), where TSSs have been detected on several 3D materials \cite{Noh2008,Hsieh2008,Chen2009a,Xia2009a,Hsieh2009}.
Among the discovered TIs, Bi$_2$Se$_3$ is of particular interest: it has a relatively large bulk band gap and a simple TSS consisting of a single Dirac conelike structure. 
This TSS is predicted to have a helical spin structure in which the spin of an electron is locked 
perpendicular to its momentum, with opposite momenta having the opposite spins. 
Spin- and angle-resolved photoemission spectroscopy (SARPES) studies have indeed confirmed the spin-momentum locked Dirac cone on the surface state of Bi$_2$Se$_3$ \cite{Hsieh2009}. However, while theories predict a complete spin polarization, experiments only show a modest spin-polarization of $\sim0.2$. This discrepancy represents a serious problem both for the fundamental understanding of TIs as well as for the prospect of their use in technological applications.
Intrinsically, a hexagonal warping of the band structure \cite{Fu2009,Chen2009a,Zhou2009} and/or a strong spin-orbit entanglement \cite{Yazyev2010} could be important factors in reducing the measured spin polarization. In addition, various scattering processes, in particular those involving the unpolarized bulk states, could also depolarize the TSS \cite{Park2010}. 
On the other hand, the apparent spin polarization could also be reduced extrinsically, due to the insufficient instrumental resolution in SARPES experiments, especially in the situation where different states are relatively close to each other. 

In this letter, we present a close comparison between the spin-resolved and high-resolution angle-resolved photoemission data from Bi$_2$Se$_3$ and we show that the lower limit on the experimentally measured spin-polarization of the TSS is $\sim 0.75$, significantly higher than in previous experiments \cite{Hsieh2009} or in recent theoretical models \cite{Yazyev2010}. Our studies indicate that the finite instrumental resolution is the dominating factor  limiting the measured polarization and point toward a full intrinsic polarization of the TSS.

ARPES studies in spin-integrated mode were performed at the beam line 12.0.1 at the Advanced Light Source. The data were recorded at $\sim15$ K using a Scienta-100 electron analyzer at photon energies ranging from 30 to 100 eV. The energy  and angle resolution were $\sim15$ meV and better than $ 0.15^{\circ}$, respectively.
The SARPES experiments were performed at the beamline U5UA at the National Synchrotron Light Source using the Omicron 
EA125 electron analyzer coupled to a mini-Mott spin polarimeter that enables measurements of the in-plane spin polarization. The spin-resolved data were recorded at 50 eV photon energy, with the sample kept at $\sim80$ K. Energy and angle resolution were approximately 45 meV and $0.5^{\circ}$, respectively. 
All the samples were cut from the same bulk piece and cleaved and measured in ultrahigh vacuum conditions (base pressure better than $5\times 10^{-9}$ Pa in both the ARPES and SARPES chambers).

\begin{figure}[htb]
\begin{center}
\includegraphics[width=7cm]{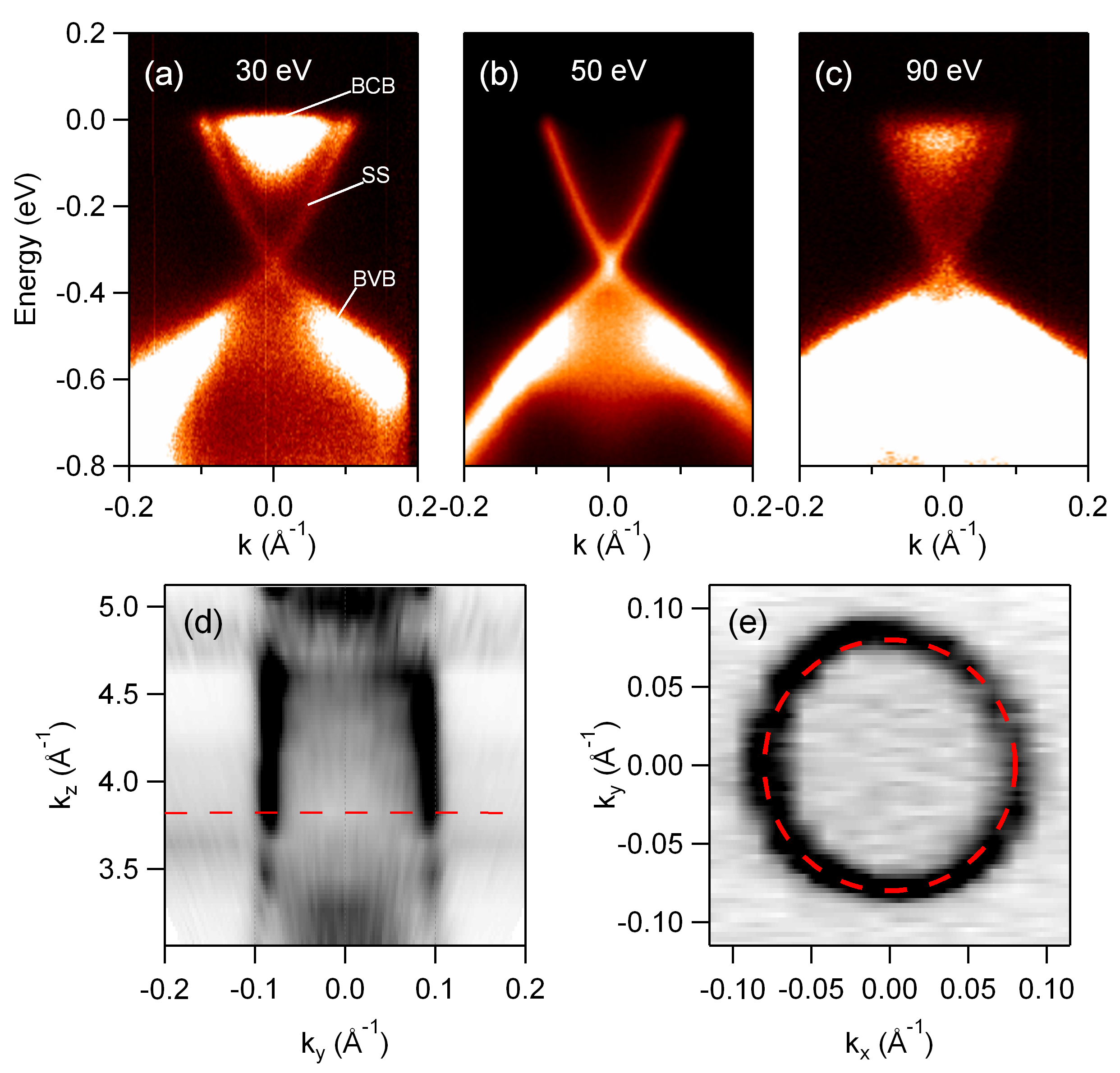}
\caption{ARPES from Bi$_2$Se$_3$ at various photon energies. (a)-(c) ARPES intensity as a function of energy and momentum recorded at 30, 50, and 90 eV photon energy. The bands are labeled in (a). (d) The Fermi surface intensity map as a function of $k_x$ and $k_z$. The data were measured at photon energies ranging from 30 to 100 eV in 5 eV steps. The nondispersive feature along $k_z$ is from the TSS. Red dashed line indicates corresponds to 50 eV photon energy. (e) The Fermi surface as a function of $k_x$ and $k_y$ measured at 50 eV photon energy. The red dashed circle  is a guide to the shape of the Fermi surface.
}
\label{Fig1}
\end{center}
\end{figure}
Figure \ref{Fig1} illustrates the photon energy dependence of the ARPES intensity from Bi$_2$Se$_3$ around the center of the surface Brillouin zone. The three upper panels show the intensity as a function of energy and momentum. The rapidly dispersing conical band represents the TSS with the Dirac point around 0.33 eV below the Fermi level. 
The bright intensity at binding energy higher than 0.4 eV is from the bulk valence bands (BVB), while the intensity near the Fermi level, inside the TSS cone, visible at 30 and 90 eV photon energy is from the bulk conducting band (BCB). Partial occupation of the BCB indicates the electron doping of Bi$_2$Se$_3$ by Se vacancies, consistent with previous reports \cite{Xia2009a}.  
However, in the spectrum measured at 50 eV this feature is completely absent and the only state crossing the Fermi level is the conical TSS. The TSS remains the same in all the spectra measured at different photon energies. The evolution of BCB with photon energy reflects the dispersion along $k_z$.
Even though $k_z$ is not conserved in a photoemission process, it can be approximated by $k_{z}=\frac{1}{\hbar}\sqrt{2m(E_{k}cos^{2}(\theta)+V)}$, where $E_k$ is the kinetic energy of a photoelectron and V is the inner potential.
A comparison with band calculations \cite{Zhang2009} suggests that the spectra measured at 30 and 90 eV correspond to a $k_z$ position close to the $\Gamma$ point in the bulk Brillouin zone, where BCB has the lowest energy, whereas the spectra measured near 50 eV correspond to a $k_z$ position close to the $Z$ point, where BCB has the highest energy. 
Since the TSS is localized to the surface, it does not disperse with $k_z$.
This can be clearly seen in Fig.\ref{Fig1}(d) where we plot the intensity at the Fermi level in $k_z$-$k_y$ plane.
The intensity map is converted into $k$ space by using the inner potential V = 10 eV.
The straight feature at $k_y = \pm 0.08$ \AA$^{-1}$ is the Fermi surface of the TSS. The BCB is visible for $k_z> 4.5 $ $\AA^{-1}$ and $k_z< 3.5$ $ \AA^{-1}$, whereas it resides above the Fermi level in between.

Figure \ref{Fig1}(e) shows the Fermi surface of Bi$_2$Se$_3$ in the $k_x$-$k_y$ plane measured at 50 eV photon energy. It consists of a single contour, well approximated by a circle. We consider this an ideal condition for measuring the spin polarization of the TSS as the BCB does not contaminate the modestly resolved SARPES picture.

To determine the $k$-position for spin-resolved measurements, we have first performed a set of spin-integrated measurements with the spin-resolved instrument. 
\begin{figure}[hb]
\begin{center}
\includegraphics[width=7cm]{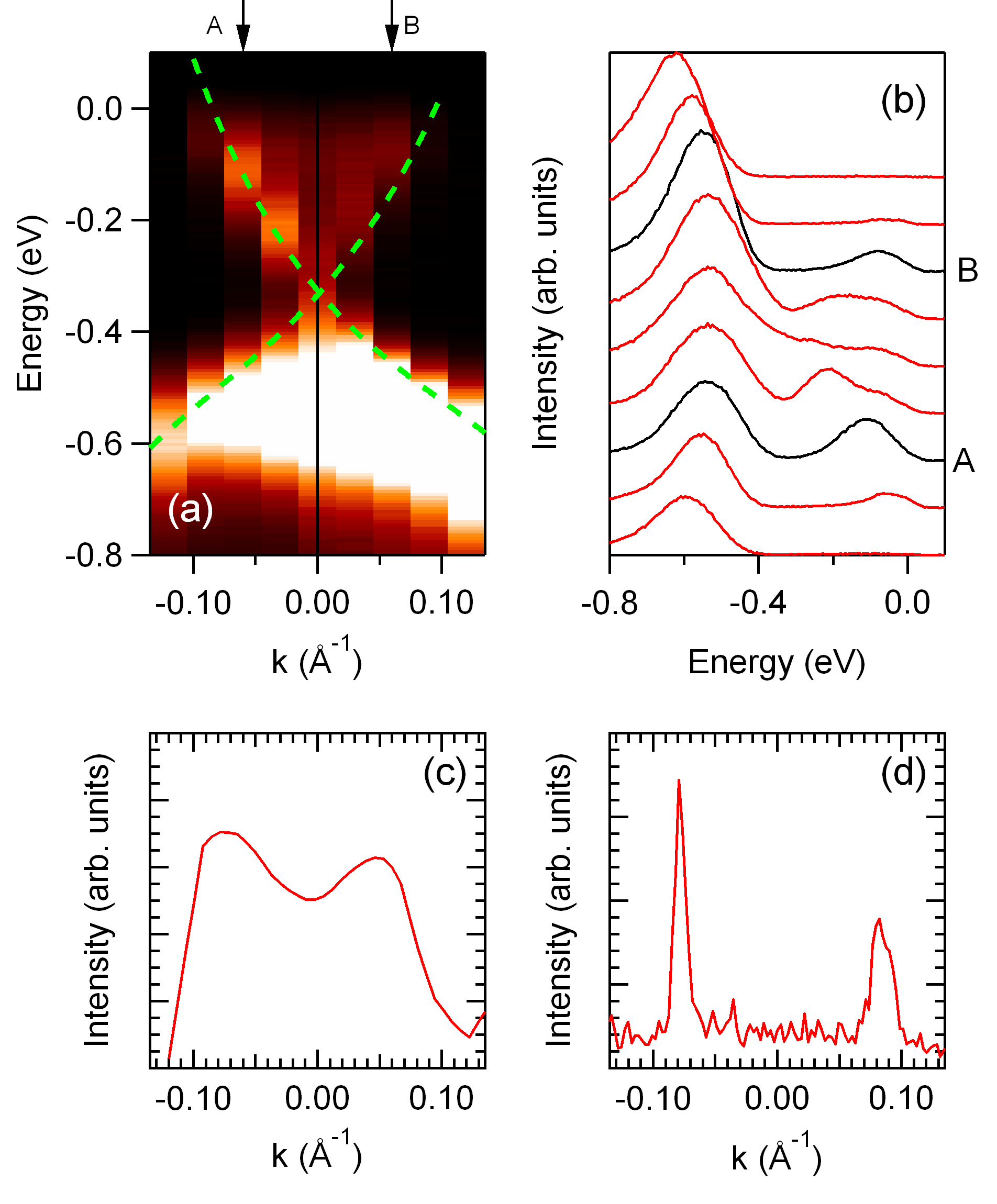}
\caption{Spin-integrated spectra of Bi$_2$Se$_3$ measured at 50 eV photon energy using the SARPES analyzer. (a) ARPES spectra along the momentum line $\Gamma M$ near the center of the surface BZ, measured in steps of $0.5^{\circ}$. The green line represents the TSS dispersion, extracted from momentum distribution curves (MDCs) from the ARPES spectrum shown in Fig.\ref{Fig1} (b). A and B indicate the $k$ positions where the spin-resolved measurements were performed. (b) energy distribution curves (EDCs) from (a). (c) MDC at the Fermi level from intensity contour in (a), interpolated to 4 times the number of original data points. (d) MDC from the Fermi surface  map in Fig.\ref{Fig1}(e) along the $k_y = 0$ line. The Lorentzian fit gives a full with at half maximum of 0.009 \AA$^{-1}$, or 0.15$^{\circ}$. 
}
\label{Fig2}
\end{center}
\end{figure}
Figure \ref{Fig2}(a) shows the 2 dimensional ARPES intensity map constructed from half a degree spaced energy distribution curves (EDCs). It is clear that both the energy and momentum resolution of the instrument were sufficient to obtain a picture of the TSS with an adequate degree of details.
The overall dispersion of the TSS is similar to the one measured in the high-resolution studies from Fig.\ref{Fig1}(b), shown here as the dashed green line, obtained by fitting the momentum distribution curves (MDC) \cite{Valla1999} with Lorenzian lineshapes.
Spin-resolved measurements were later performed at points A and B, approximately on the opposite sides of the TSS cone, slightly inside the Fermi surface to avoid the Fermi function cutoff. 
Fig.\ref{Fig2}(b) shows the corresponding spin-integrated EDCs. The two black EDCs correspond to points A and B.
These spectra have a relatively simple structure with two well separated peaks: one at $\approx -0.1$ eV corresponding to the TSS and the other one around -0.55 eV, corresponding to the BVB.

\begin{figure}[htb]
\begin{center}
\includegraphics[width=8.5cm]{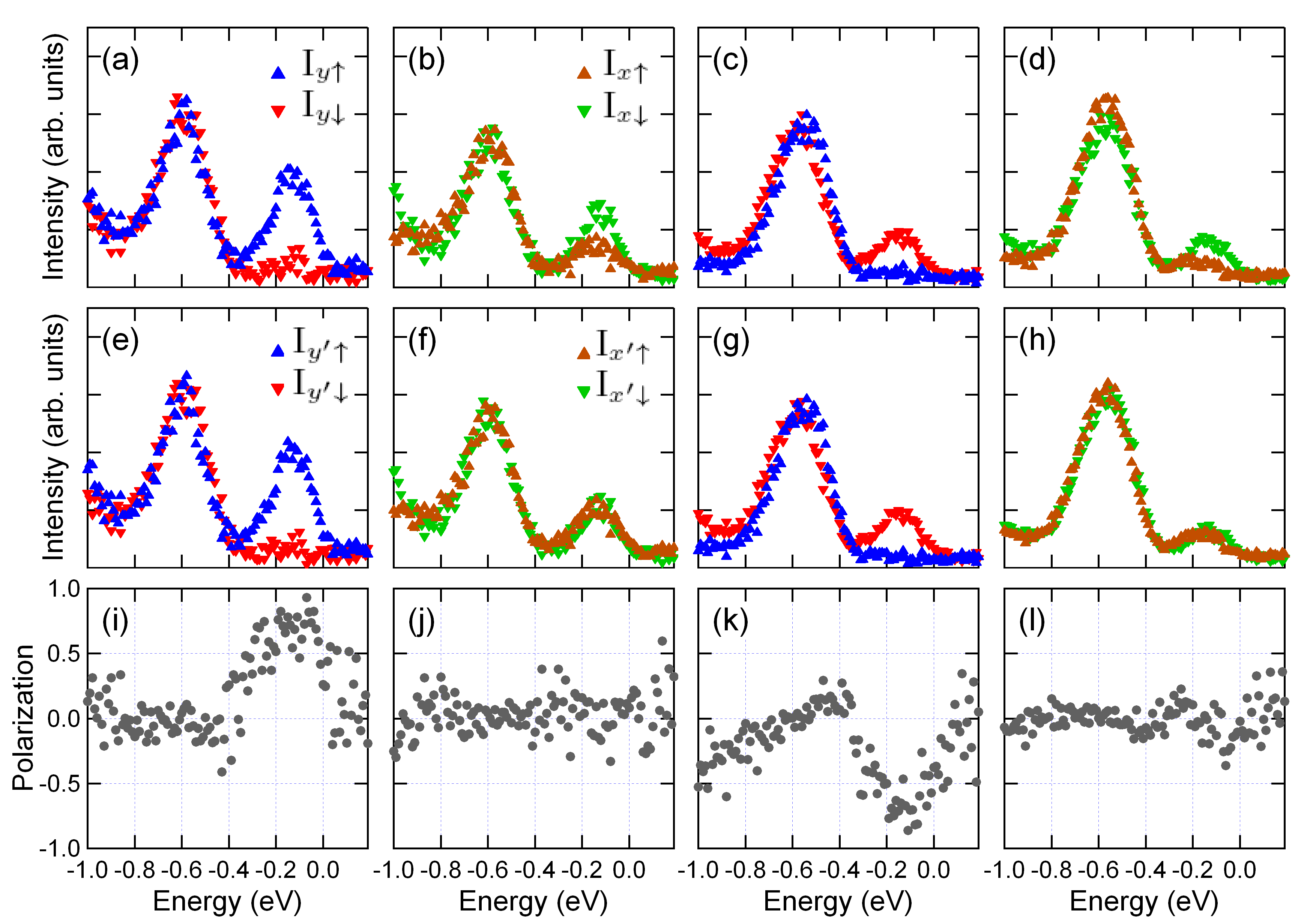}
\caption{The spin-resolved photoemission of the TSS of Bi$_2$Se$_3$.  (a-b) Spin-resolved spectra with the polarization vector along $y$ and $x$ direction for point A from Fig. \ref{Fig2}(a). (c-d) same for point B. (e-h) Spin-resolved spectra for points A and B with the polarization vectors along the red arrows in Fig.\ref{Fig4}. The corresponding spin polarizations are shown in (i-l) 
All the spectra were measured at photon energy 50 eV at T = 77 K. 
}
\label{Fig3}
\end{center}
\end{figure}

The spin-resolved results are shown in Fig. \ref{Fig3}. 
Four detectors (Left, Right, Top and Bottom) in the Mott polarimeter are oriented so that the asymmetry in Left  - Right  intensities, $A=\frac{I_L-I_R}{I_L+I_R}$, reflects the spin polarization in $y$ direction, while the asymmetry in the Top-Bottom intensities reflects the polarization in $x$ direction \cite{Gay1992}. The spin polarization is obtained as $P=\frac{A}{S}$ where $S$ represents the Sherman function, or the efficiency of the spin polarimeter, $S = 0.17$ in our case.

Figure \ref{Fig3}(a) shows the spin-resolved spectra with polarization defined in the $y$ direction for point A from Fig. \ref{Fig2}.
The peak at $\sim -0.1$ eV, corresponding to the TSS, has almost pure spin-up character, while the spectrum appears essentially unpolarized below -0.5 eV. This suggests that the bulk states are not polarized and that the TSS is almost completely polarized in the $y$ direction.
In the $x$ direction (Fig.\ref{Fig3}(b)), the TSS peak shows only a weak polarization with the spin-down character.
The spin-resolved data for point B, nearly symmetric to point A relative to the $k_x = 0$ plane, are shown in panels (c) and (d). Clearly, the TSS peak shows the reversal of spin polarization in the $y$ direction when the momentum is reversed ($k_x \rightarrow -k_x$), while the small spin-polarization in the $x$ direction remains unaffected. 
We note that if the spin polarization is indeed perpendicular to the momentum, as the spin-momentum locking dictates, one would expect zero polarization in the $x$ direction if A and B points are exactly on the $k_y$=0 line.
Experimentally, however, it is very difficult to align the sample precisely along the $k_y$ = 0 line. The measured nonzero polarization in the $x$ direction suggests that points A and B have a finite of $k_y$, caused by a small tilt of the sample. In order to estimate the true degree of spin polarization of the TSS it is therefore necessary to recalculate the polarization along the axis parallel and perpendicular to the maximum and minimum spin-polarization direction. This is done in Figs. \ref{Fig3}(e)-(h) that show spin-resolved spectra and (i)-(l) showing the corresponding spin-polarizations. At point A, a $\sim 20^{\circ}$ rotation off the positive $y$ direction is sufficient to completely eliminate the residual polarization in the new $x'$ direction. Similarly, at point B, a rotation of $\sim 25^{\circ}$ off the negative $y$ direction is necessary. There is a hint in the spin-resolved spectra that the low binding energy tail of the peak at -0.55 eV might in fact be polarized and that the polarization is opposite to the polarization of the low binding energy peak. This suggests that in the energy region immediately below the Dirac point, the BVB probably overlaps with the TSS component whose polarization is reversed from the one measured above the Dirac point.
\begin{figure}[htb]
\begin{center}
\includegraphics[width=6.75cm]{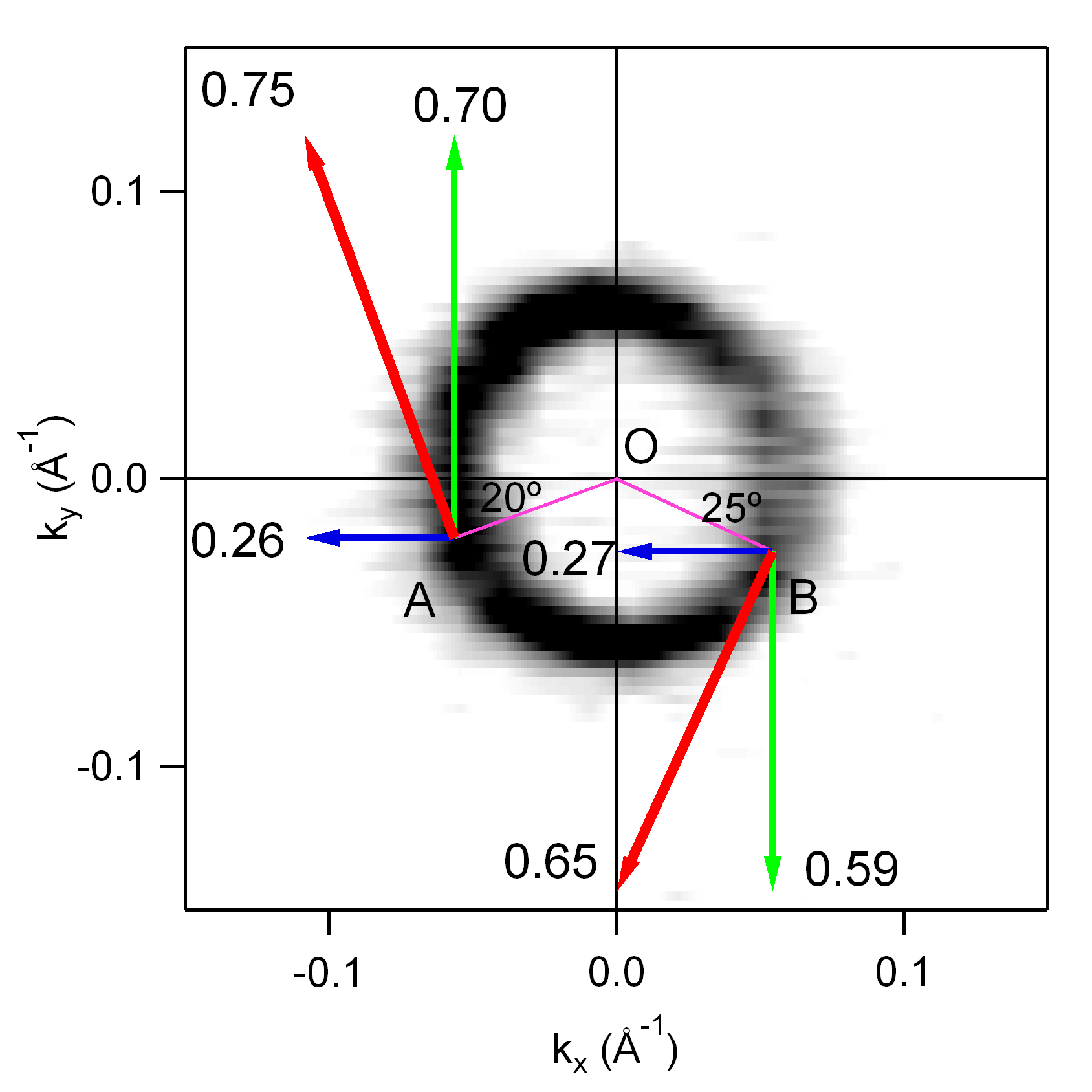}
\caption{Vector analysis of spin polarization in Bi$_2$Se$_3$. The green (blue) arrows represent the spin polarization in the $y$ ($x$) direction, determined from Fig. \ref{Fig3}. The red arrows correspond to the total spin polarization. The lengths of the arrows indicate the magnitudes of spin polarizations. The origins of the arrows correspond to the $k$- positions of points A and B. The intensity map represents the constant energy contour at -0.1 eV from spin-integrated ARPES data. 
}
\label{Fig4}
\end{center}
\end{figure}

From the measured polarizations and by assuming the spin-momentum locking, we can also determine the $k$-positions of points A and B. 
Figure \ref{Fig4} sumarizes the obtained spin-resolved results. To illustrate the $k$ positions of A and B, we also plot the constant energy contour of intensity at -100 meV from the high-resolution spin-integrated measurements.
The green (blue) arrows show the spin polarization along the $y$ ($x$) direction. The red arrows show the total in-plane spin polarization. The lengths of the arrows indicate the magnitudes of the spin polarizations, and the origins indicate the $k$ positions of points A and B, reflecting the sample tilt of $\approx 0.4^{\circ}$ from the $k_y=0$ line.
From the resulting configuration it is clear why points A and B have the opposite spin polarization in the $y$ direction and the parallel in the $x$ direction. 

The total spin-polarization of the TSS measured here approaches unity. The key factor for this observation was the appropriate choice of photon energy at which the BCB was removed form the region of (E,$\bf{k}$) space occupied by the TSS. In this way the overlap of TSS with unpolarized bulk states, which can significantly reduce the apparent polarization, was effectively avoided. An illustration of a dramatic reduction in the measured polarization is seen at energies below the Dirac point, where the TSS overlaps with the BVB. Even though the TSS is expected to be fully polarized, with reversed spin helicity, the measured polarization has almost completely diminished due to the overlap with BVB.   

We note that although the spin polarization is much higher than previously measured, it is still less than 1. One of the possible reasons for this reduction could be a small hexagonal warping of the band structure that may cause an out-of-plane canting of the spins and a reduction of the in-plane component. However, as can be seen in Fig.\ref{Fig4}, at -100 meV, where the polarization was measured, such warping is beyond our detection limits and probably has a negligible effect.
Secondly, as suggested in recent calculations \cite{Yazyev2010}, a strong spin-orbit entanglement could reduce the spin polarization down to 0.5. Our results show higher polarization, suggesting that the proposed entanglement has a much smaller effect, if any. The third reason could be the interband scattering that would mix the TSS with the unpolarized BCB. These processes should also be suppressed by our choice of $k_z$ as they would necessarily involve a finite exchanged momentum in the $z$ direction, $q_z$.
  
Finally, we argue that the measured polarization is still limited primarily by the finite angular resolution of the instrument. 
As can be seen from Fig. \ref{Fig4}, the constant energy contour at -100 meV has a very small radius, $\approx 0.06$ \AA$^{-1}$ or approximately $1^{\circ}$ in angle, just about 2 times larger than our estimated resolution in the spin-resolved mode. Therefore, there is still some overlap of the two peaks [see Fig.\ref{Fig2}(c)] with the opposite spin orientations that reduces the measured polarization.
If the SARPES could be performed with the momentum resolution comparable to the one in high-resolution ARPES [see Fig. \ref{Fig2}(d)], one would be able to measure the intrinsic spin polarization  of the TSSs. 

In summary, we measured the spin texture of the TSS on Bi$_2$Se$_3$ in high-resolution spin-resolved photoemission. We found that when the TSS is well separated from the bulk states, its measured spin-polarization approaches unity. Our results have important implications for both the fundamental research and potential applications of TIs. 

We acknowledge useful discussions with M. Khodas and P. D. Johnson. The work at Brookhaven is supported by the US Department of Energy (DOE) under Contract No. DE-AC02-
98CH10886. The work at MIT is supported by the DOE under Grant No. DE-FG02-04ER46134.
ALS is operated by the US DOE under
Contract No. DE-AC03-76SF00098.

\providecommand{\noopsort}[1]{}\providecommand{\singleletter}[1]{#1}%

\end{document}